\documentclass[a4paper,11pt]{article}
\pdfoutput=1 
\usepackage{jheppub}
\usepackage{graphicx}
\usepackage{epstopdf}
\usepackage{dcolumn}
\usepackage{xcolor}
\usepackage{amsmath,amssymb}
\usepackage{hyperref}
\usepackage[utf8]{inputenc}
\usepackage[english]{babel}
\usepackage[displaymath, mathlines]{lineno}
\usepackage{blindtext}
\usepackage{ifthen}
\usepackage{orcidlink}
\usepackage{physics}
\usepackage{comment}

\graphicspath{{figures/}}
\newboolean{articletitles}
\setboolean{articletitles}{true}

\input{belle2-symbols}

\newcommand{\kekpreprint}{2026-8}  
\newcommand{\preprintbelle}{2026-012}  

\begin{document}

\includegraphics[width=3cm]{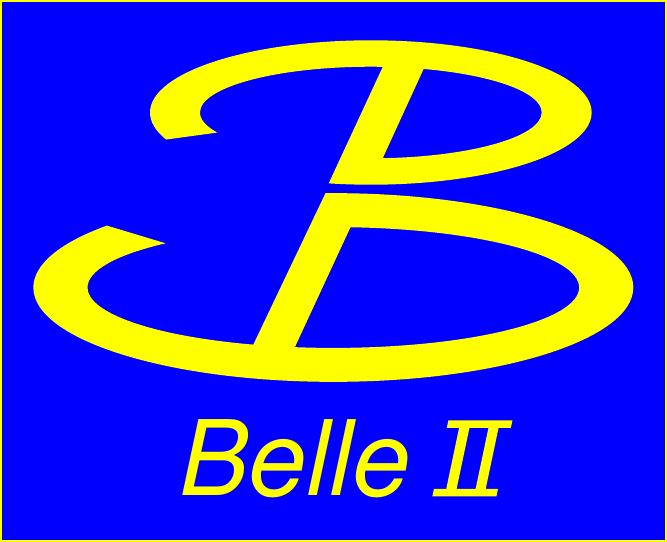}\vspace*{-3cm}
\begin{flushright}

KEK preprint: \kekpreprint \\
Belle II preprint: \preprintbelle \\

\end{flushright}
\vspace*{44pt}
\title{Search for the lepton-flavor-violating decay $ \tau^{\pm} \to \mu^{\pm} \gamma$ at Belle II
}
\collaboration{The Belle II Collaboration}
  \author{M.~Abumusabh\,\orcidlink{0009-0004-1031-5425},} 
  \author{I.~Adachi\,\orcidlink{0000-0003-2287-0173},} 
  \author{A.~Aggarwal\,\orcidlink{0000-0002-5623-3896},} 
  \author{H.~Ahmed\,\orcidlink{0000-0003-3976-7498},} 
  \author{Y.~Ahn\,\orcidlink{0000-0001-6820-0576},} 
  \author{H.~Aihara\,\orcidlink{0000-0002-1907-5964},} 
  \author{M.~Akdag\,\orcidlink{0009-0004-3728-1077},} 
  \author{N.~Akopov\,\orcidlink{0000-0002-4425-2096},} 
  \author{S.~Alghamdi\,\orcidlink{0000-0001-7609-112X},} 
  \author{M.~Alhakami\,\orcidlink{0000-0002-2234-8628},} 
% ...

\emailAdd{coll-publications@belle2.org} 
\abstract{
We present a search for the lepton-flavor-violating decay $\tau^{\pm}\to\mu^{\pm}\gamma$ using a data sample that corresponds to an integrated luminosity of 428 fb$^{-1}$ recorded by the Belle II experiment at the SuperKEKB asymmetric-energy $e^{+}e^{-}$ collider. We employ a multivariate classifier to suppress the backgrounds from the Standard Model processes, and the signal extraction is performed using an extended maximum-likelihood fit. Since no significant excess over the expected background is observed, we set an upper limit on the branching fraction  $\mathcal{B}(\tau^{\pm}\to\mu^{\pm}\gamma) <  9.5$ $ (12.2)\times10^{-8}$ at the 90\% (95\%) confidence level, using the CL${_s}$ technique.
}  

\maketitle
\flushbottom 

\section{Introduction}

Charged lepton flavor violation is suppressed in the Standard Model~(SM) even when small but finite neutrino masses are taken into account. The branching fractions of charged-lepton-flavor-violating processes are unobservably small, of the order of $\sim 10^{-54}$ \cite{Calibbi:2017uvl,Petcov:1976ff,PhysRevD.16.1444}. Therefore, any observation of these processes would provide clear evidence of physics beyond the SM.

Moreover, several SM extensions, such as supersymmetric models, supersymmetric models with a see-saw mechanism, grand unified theories, the Littlest Higgs models, etc., predict these processes at experimentally accessible rates. The radiative decay $\tau^{\pm}\rightarrow\mu^{\pm}\gamma$ is predicted to have a branching fraction of the order of 10$^{-10}-10^{-8}$ in these new physics theories~\cite{Blanke_2007, Brignole_2003, Fukuyama_2003, Masiero_2003}, constituting a key channel for both discovering charged lepton flavor violation and testing the SM extensions.

Several searches for $\tau^{-}\rightarrow\mu^{-}\gamma$ decays\footnote{The inclusion of the charge-conjugate decay mode is implied throughout this paper.} have been conducted by the Belle, BaBar, and CLEO experiments~\cite{Uno2021,Babar2009, Cleo2000} over the past two decades. No evidence of a signal has been found. The most stringent upper limit (UL) on the branching fraction has been reported by the Belle experiment, $\mathcal{B}(\tau^{-}\rightarrow\mu^{-}\gamma)<4.2\times10^{-8} $ at the 90\% confidence level (C.L.), using its full data set corresponding to an integrated luminosity of 988.0~fb$^{-1}$~\cite{Uno2021}. The BaBar experiment had set an UL on the branching fraction, $\mathcal{B}(\tau^{-}\rightarrow\mu^{-}\gamma)<4.4\times10^{-8} $ at the 90\% C.L. based on a data sample of 515.5~fb$^{-1}$ of integrated luminosity~\cite{Babar2009}. The CLEO experiment obtained an UL of $\mathcal{B}(\tau^{-}\rightarrow\mu^{-}\gamma)<1.1\times10^{-6} $ at the 90\% C.L. with 13.8~fb$^{-1}$ of integrated luminosity~\cite{Cleo2000}.

This paper presents a search for the lepton-flavor-violating decay $\tau^{-}\rightarrow\mu^{-}\gamma$ using data from the Belle II experiment. To maximize sensitivity, we employ a multivariate classifier to suppress background and enhance signal detection efficiency. The signal yield extraction is performed using a two-dimensional fitting procedure, having modeled both the signal and background components. Background and signal shapes are fixed to the values extracted from fits to simulated samples, allowing the yields to float freely in the fit to the data. We then construct a test statistic with an unbinned extended likelihood function that incorporates systematic uncertainties as nuisance parameters. The UL is computed using the CL$_{s}$ technique \cite{cls_method, Junk_1999}, which provides a conservative limit in the case of a downward fluctuation of the background yield.

In this analysis, simulated signal and background samples are used to determine the signal efficiency and fix the selection criteria, respectively, while sideband data are used to confirm that the simulation accurately represents the data and to estimate background composition. The signal region is kept hidden until the selection criteria and background estimations are finalized to avoid any experimenter bias.

The subsequent sections are structured as follows. Section \ref{section2} describes the Belle~II detector along with the simulated and experimental samples. Section \ref{section3} presents the event selection and Section \ref{sectionbkg} the background suppression. Sections \ref{section4} and \ref{section5} discuss the signal extraction model and the treatment of systematic uncertainties. Finally, Sections \ref{section6} and \ref{section7} present the results and conclusions, respectively.
\section{Belle II detector and Datasets}\label{section2}
The Belle~II detector is located at the SuperKEKB asymmetric-energy accelerator, which collides 7 GeV electrons and 4 GeV positrons at center-of-mass (c.m.) energies near the $\Upsilon(4S)$ resonance~\cite{Akai:2018mbz}. The Belle II detector~\cite{Abe:2010gxa} has a cylindrical geometry and includes a 2-layer silicon-pixel detector~(PXD) surrounded by a 4-layer
double-sided silicon-strip detector~(SVD)~\cite{Belle-IISVD:2022upf} and a 56-layer central drift chamber~(CDC). These detectors reconstruct trajectories of charged particles (tracks).  Only one-sixth of the second
layer of the PXD was installed for the data analyzed here. The $z$-axis of the Belle II detector is defined along the electron beam direction; the transverse plane is the plane perpendicular to this axis. The detector is divided into forward endcap, barrel, and backward endcap regions according to the polar angle $\theta$ with respect to the $z$-axis, corresponding approximately to $\theta < 32^{\circ}$, $32^{\circ} < \theta < 130^{\circ}$, and $\theta > 130^{\circ}$, respectively. Surrounding the
CDC, which also provides $dE/d{x}$ energy-loss measurements, is a time-of-propagation counter~(TOP)~\cite{Kotchetkov:2018qzw} in the central region, and an aerogel-based ring-imaging Cherenkov counter~(ARICH) in the forward region.  These detectors provide charged-particle identification.  Surrounding the TOP and ARICH is an electromagnetic calorimeter~(ECL) based on CsI(Tl) crystals that primarily provides energy and timing measurements for photons and electrons. Outside of the ECL is a superconducting solenoid magnet that provides a 1.5~T magnetic field that is parallel to the $z$-axis. Its flux return is instrumented with resistive-plate chambers
and plastic scintillator modules to detect muons, $K^0_L$ mesons, and neutrons~\cite{KETTER2026170893}.


We use Monte Carlo~(MC) simulated samples at the $\Upsilon(4S)$ energy to optimize the event selection, to estimate the signal detection efficiency, and to model the shape of the signal and backgrounds.
The $e^{+}e^{-}\to\tau^{+}\tau^{-}$ events are generated using the KKMC generator~\cite{Jadach:1999vf}. Signal samples are simulated as $e^{+}e^{-}\to\tau^{+}\tau^{-}$ events with one of the $\tau$ decaying to the $\mu^{\pm}\gamma$ final state and the other $\tau$ decaying according to known branching fractions~\cite{ParticleDataGroup:2022pth}. Background processes include $\epem \to \qqbar$ events, where $q$ indicates a $u$, $d$, $s$, or $c$ quark;  $\epem \to \Upsilon(4S) \to B\Bar{B}$ events, where $B$ denotes a charged or neutral $B$ meson; $\epem\to \ell^+ \ell^-$, where $\ell$ indicates an electron or a muon; $\epem \to \ell^+\ell^-h^+h^-$ events, where $h$ indicates a pion, kaon, or proton; four-lepton processes $\epem \to e^+e^-e^+e^-$, $\mu^+\mu^-\mu^+\mu^-$, $\mu^+\mu^-e^+e^-$, $e^+e^-\tau^+\tau^-$, $\mu^+\mu^-\tau^+\tau^-$ and $\tau^+\tau^-\tau^+\tau^-$; and $e^+e^- \to \gamma \hspace{1mm}+$ hadrons, where the photon emission is caused by initial-state radiation (ISR) and hadrons include $K^+K^-$, $ K^0\bar{K}^0$, $ \pi^+\pi^-$, and $\pi^+\pi^-\pi^{0}$ combinations. The $\tau$ decays are simulated by the TAUOLA generator~\cite{Tauola,chrzaszcz2017tauolatauleptondecays,ANTROPOV2023108592}, with ISR and final-state radiation~(FSR) from the tau production vertex modeled by KKMC, while photon radiation in $\tau$ decays is modeled by the PHOTOS package~\cite{BARBERIO1994291}.
The $e^{+}e^{-}\to\mu^{+}\mu^{-}$ and $e^{+}e^{-}\to q\bar{q}$ background processes are simulated using KKMC. Fragmentation of $q\bar{q}$ pairs is simulated using the PYTHIA package~\cite{Sjostrand:2014zea}.
The $e^{+}e^{-}\to B\Bar{B}$ background is simulated using the EvtGen package~\cite{Lange:2001uf}, interfaced to PYTHIA. The EvtGen and KKMC generators also use PHOTOS to simulate FSR. 
The background $e^{+}e^{-}\to e^{+}e^{-}$ is simulated using the BabaYaga@NLO~\cite{BALOSSINI2006227, BALOSSINI2008209, CARLONICALAME200448, CARLONICALAME200116,CARLONICALAME2000459} generator. The AAFH~\cite{BERENDS1985421, BERENDS1985441, BERENDS1986285} and TREPS \cite{uehara2013trepsmontecarloeventgenerator} generators are used for the production of the four-lepton and $e^+e^- \to \ell^{+}\ell^{-}h^+h^-$ processes. The $e^+e^- \to \gamma \hspace{1mm}+$ hadrons backgrounds are produced using the PHOKHARA generator \cite{Rodrigo2002}. The simulation of the detector response is done by the Geant4 software package \cite{AGOSTINELLI2003250}.

The analysis uses a data sample collected from 2019 to 2022 corresponding to an integrated luminosity of 428 $\pm$ 2~fb$^{-1}$ \cite{luminosity_belle2}. The dataset is collected at the $e^{+}e^{-}$ c.m. energies of 10.58 GeV~(365.4 fb$^{-1}$), 10.52 GeV~(42.7 fb$^{-1}$) and at various energies around 10.75 GeV~(19.8 fb$^{-1}$), and corresponds to $N_{\tau\tau}=393\times10^{6}$ events. 
The total integrated luminosity of the simulated $e^{+}e^{-}\to\tau^{+}\tau^{-}$ and $\epem \to \qqbar$ samples is 9.5 ab$^{-1}$. For the remaining processes, the equivalent integrated luminosity is 2.4 ab$^{-1}$, except for $\epem \to e^+e^-e^+e^-$, $\mu^+\mu^-\mu^+\mu^-$ for which 560 fb$^{-1}$ are used, and $\epem \to e^+e^-$ for which 140 fb$^{-1}$ is used.

The data are selected by a hardware trigger based on energy deposits (clusters) and their topologies in the ECL, with a retention rate of about 96\%, measured with simulated signal samples. The main selection requirement is a total ECL energy greater than 1 GeV and a topology incompatible with Bhabha events. The Belle II analysis software is used to process all samples~\cite{basf2-zenodo}.

\section{Event Selection}\label{section3}

Events are required to have exactly two oppositely charged particles originating from the interaction point to suppress $e^{+}e^{-}\to q\bar{q}$ and $e^{+}e^{-}\to B\bar{B}$ events. We select tracks with a distance of closest approach smaller than 1.0~cm in the transverse plane and smaller than 3.0~cm along the $z$-axis. 
Muon candidates are identified using a likelihood-based discriminator $P_{\mu}=L_{\mu}/(L_e+L_\mu+L_\pi+L_K+L_p+L_d)$, where the likelihoods $L_{j}$ for each charged-particle hypothesis~(electron, muon, pion, kaon, proton, and deuteron, respectively) combine particle-identification information from all detectors except the PXD and SVD. We require $P_{\mu}>0.95$ to select muons. The signal muon is required to have a momentum magnitude of $p_{\mu}>1.0$~GeV/$c$. Photons are reconstructed from energy deposits in the ECL that are not associated with a track. They are required to have an energy greater than 200 MeV, a polar angle in the laboratory frame satisfying $-0.866<\cos\theta_{\gamma}<0.957$, a cluster timing within 200 ns of the event time, and the absolute value of the cluster timing over its uncertainty must be less than two. The event time is calculated using the timing information of reconstructed charged particles. These requirements effectively suppress photons arising from beam-related backgrounds. Neutral pions are reconstructed from pairs of photons satisfying the above selection but with a relaxed energy requirement of $E_{\gamma}>100$ MeV and invariant mass of 115~MeV/$c^{2}<m_{\gamma\gamma}<152$~MeV/$c^{2}$.

In the $\mathrm{c.m.}$ frame, the $e^{+}e^{-}\to\tau^{+}\tau^{-}$ events produce two $\tau$ leptons with back-to-back momenta; thus, the decay products of each $\tau$ are isolated from the others and contained in opposite hemispheres. To separate the event into two hemispheres, we use the thrust axis, $\hat{t}$, which is the unit vector that maximizes the thrust magnitude:
\begin{equation}
    V_{\mathrm{thrust}}=  \max_{\hat{t}} \left(  \frac{ \sum_{i}\mid \vec{p}_{i}^{\hspace{0.6mm}*} \cdot \hat{t}\mid}{\sum_{i} \mid\vec{p}_{i}^{\hspace{0.6mm} *} \mid} \right)\;,
\label{eq:thrust}
\end{equation}
where $\vec{p}_{i}^{\hspace{0.6mm} *}$ is the $\mathrm{c.m.}$ momentum of each reconstructed particle.\footnote{Throughout this paper, quantities evaluated in the c.m. frame are denoted with a superscript asterisk.} The distribution of the thrust observable is presented in the upper panel of Figure \ref{fig:xi_distribution}.

We define the signal-side hemisphere as that containing the $\tau^{-}\to\mu^{-}\gamma$ candidate, and the tag-side hemisphere as the opposite hemisphere, which contains the decay products of the other $\tau$ lepton. The tag side is required to contain a single charged particle with a muon veto of $P_{\mu}<0.1$. This criterion retains a broad range of $\tau^{+}$ decay modes, including $\tau^{+}\to e^{+}\nu_{\ell}\bar{\nu}_{\tau}$, $\tau^{+}\to\pi^{+}\bar{\nu}_{\tau}$, and $\tau^{+}\to\rho^{+}\bar{\nu}_{\tau}$. The muon veto on the tag side suppresses the dominant $e^{+}e^{-}\to\mu^{+}\mu^{-}$ background, which mimics our signal through the presence of an ISR photon.

Based on the thrust axis, each reconstructed particle is therefore assigned to either the signal or tag hemisphere. We require events with exactly one photon on the signal side, reconstructed as detailed above, but with an increased energy threshold of $E_{\gamma} > 1.0$ GeV. Events with additional photons with energy larger than 200 MeV on the signal side are discarded.

Finally, we select events with a magnitude of the missing momentum $p_{\mathrm{miss}} > 0.5$~GeV/$c$, where $p_{\mathrm{miss}}$ is defined as the difference between the beam momenta and the sum of the laboratory‑frame momenta of all reconstructed particles. 
This requirement helps suppress hadronic final states produced in two-photon interactions, $e^+e^-\to e^+e^-Nh$ with $N>2$, which are not included in the simulated samples and tend to populate the low missing-momentum region.

After applying the baseline selection described above, the signal efficiency is 13.8\%. The background composition is dominated by $e^{+}e^{-}\to\tau^{+}\tau^{-}$  events (96.0\%), followed by $e^{+}e^{-}\to\mu^{+}\mu^{-}$ (2.0\%), $e^{+}e^{-}\to q\bar{q}$ (1.0\%) and other processes (1.0\%). These values and subsequent results are obtained after applying correction factors to the simulation to account for differences between data and simulation, including lepton identification, photon reconstruction efficiency, and trigger efficiency.

The signal is extracted using the beam-constrained mass $M_{\mathrm{bc}}$ and the normalized energy difference $\Delta E / \sqrt{s}$ kinematic variables. The beam-constrained mass is defined as
\begin{equation}\label{eq:mbc}
    M_{\mathrm{bc}} = \sqrt{(E^{*}_{\mathrm{beam}})^{2} - (|\vec{p}^{\hspace{0.6mm}*}_{\mu \gamma}| )^{2}}\;,
\end{equation}
where $|\vec{p}^{\hspace{0.6mm}*}_{\mu \gamma}|$ is the muon and photon system momentum magnitude in the $\mathrm{c.m.}$ frame, and $E^{*}_{\mathrm{beam}}=\sqrt{s}/2$ is the beam energy with $\sqrt{s}$ the $\mathrm{c.m.}$ energy. To compute the right term of Eq.~(\ref{eq:mbc}), we assume $|\vec{p}_{\gamma}^{\hspace{0.6mm}*}|=E_{\gamma}^{*} = E^{*}_{\mathrm{beam}} -E^{*}_{\mu}$. The normalized energy difference is defined as
\begin{equation}\label{eq:deltae}
 \Delta E / \sqrt{s} =(E_{\mu \gamma}^{*}-E_{\mathrm{beam}}^{*} )/ \sqrt{s}\;,
\end{equation}
where $E_{\mu \gamma}^{*}$ is the energy of the muon and photon system in the $\mathrm{c.m.}$ frame; for the computation, $E_{\gamma}^{*}=|\vec{p}_{\gamma}^{\hspace{0.6mm}*}|$ is used as the deposited energy in the ECL. The signal events peak at $M_{\mathrm{bc}}$ values around $m_{\tau}$ and $\Delta E$ around zero. Radiative effects from ISR and FSR cause a spread around these nominal values, producing tails in the negative-$\Delta E$ and large-$M_{\mathrm{bc}}$ regions of the distributions. Figure \ref{fig:mc_data_signalregion} shows the signal distribution after all selections.

The signal region, defined as $M_{\mathrm{bc}} \in$~$[1.736, 1.819]$~GeV/$c^{2} \times \Delta E / \sqrt{s} \in  [-0.02, 0.02]$, is kept hidden until the selection criteria and background estimation were finalized. The sidebands are defined as data outside the signal region but falling within the $M_{\mathrm{bc}} \in [1.6, 2.0]$ GeV/$c^{2} \times \Delta E / \sqrt{s} \in [-0.02, 0.02]$ region.

\section{Background Suppression}\label{sectionbkg}
A gradient-boosted decision tree (GBDT)~\cite{bib:xgb} is used to suppress the background events that pass the previous selections.
The classifier is trained on simulated samples of the dominant backgrounds $e^{+}e^{-}\to\tau^{+} \tau^{-}$ and $e^{+}e^{-}\to\mu^{+}\mu^{-}$ along with signal events $e^{+}e^{-}\to\tau^{+} \tau^{-}[\to\mu^{-}\gamma]$. The main component of the $e^{+}e^{-}\to\tau^{+}\tau^{-}$ background comes from the decay $\tau^{-} \to \mu^{-} \bar{\nu}_{\mu} \nu_{\tau}$, where the muon can be reconstructed together with an ISR photon or a beam-background photon.

The GBDT is trained inclusively on all possible tag-side modes on a set of global-event and kinematic variables that provide discrimination between signal and background processes based on their distinct kinematic topologies. For the GBDT training, we exclude input variables found to be highly correlated, with a Pearson correlation larger than 0.5, with either $M_{\mathrm{bc}}$ or $\Delta E / \sqrt{s}$ to prevent the introduction of bias through indirect reconstruction of the signal region.

\begin{figure}[h!]
    \centering
    \includegraphics[width=0.7\linewidth]{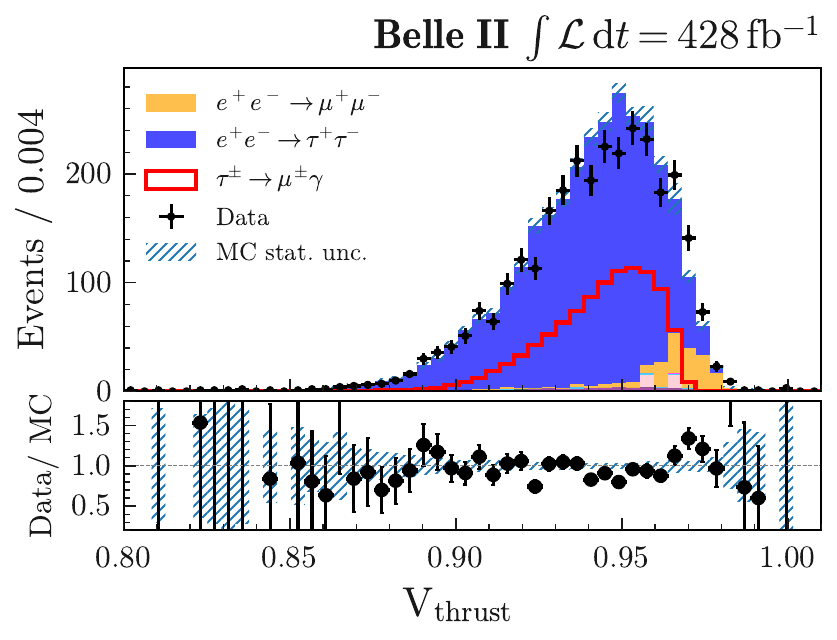}
    \par
    \includegraphics[width=0.7\linewidth]{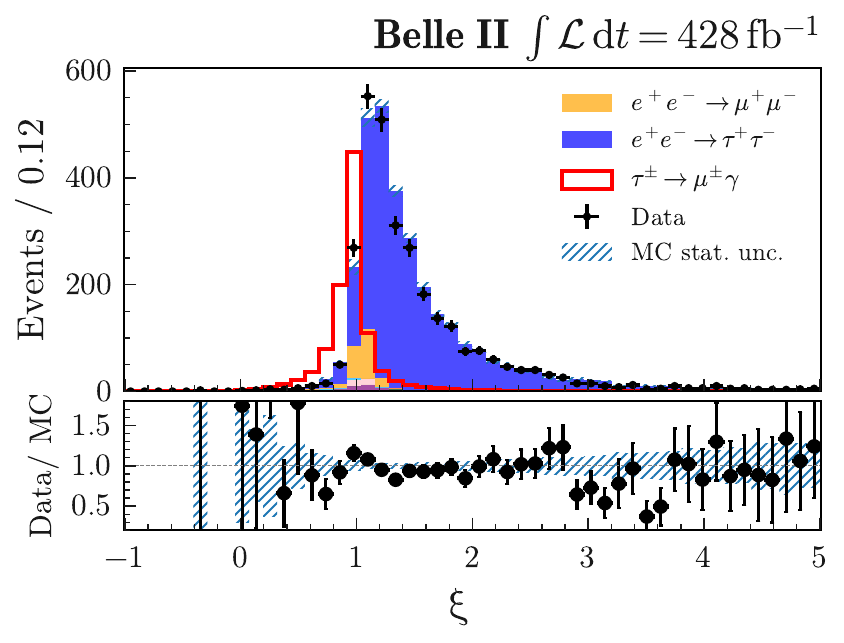}
    \caption{Distributions of the thrust (upper panel) and $\xi$ (lower panel) observables after applying the baseline selection (prior to the GBDT cut) in the sideband region. Simulated background processes are stacked and normalized to an integrated luminosity of 428~fb$^{-1}$, with only the dominant contributions indicated in the legend. The $\tau^{-} \to \mu^{-} \gamma$ signal simulation is overlaid with an arbitrary normalization factor for illustration. The lower panel in each plot displays the ratio of data to the total simulated background.}\label{fig:xi_distribution}
\end{figure}

The global-event variables include the thrust magnitude, which is distributed at higher values for the $e^{+}e^{-}\to\mu^{+}\mu^{-}$ background than for other processes (see Figure~\ref{fig:xi_distribution}); the squared mass of the unreconstructed neutrino system $m_{\nu}^{2}=\left( E^{*}_{\mu \gamma}-E^{*}_{\mathrm{tag}}\right)^{2} -\left(p^{*}_{\mathrm{miss}} \right)^{2}$, where $E^{*}_{\mathrm{tag}}$ is the sum of the tag side energy in the $\mathrm{c.m.}$ frame; $m_{\nu}^{2}$ tends to be positive for signal events, while the background exhibits a tail extending to negative values; and the missing momentum magnitude $p_{\mathrm{miss}}$, which helps suppress the $e^{+}e^{-}\to q\bar{q}$ and $e^{+}e^{-}\to\mu^{+}\mu^{-}$ backgrounds, as these processes populate the low missing momentum magnitude region while signal is almost uniformly distributed. Additional global-event variables include both the number of photons and the number of neutral pions in the tag-side hemisphere.

We use the following kinematic variables to distinguish signal from background: the tag-side charged particle $\mathrm{c.m.}$ momentum, in which the $e^{+}e^{-}\to\mu^{+}\mu^{-}$ background peaks at higher values than the signal; the energy of the signal-side photon, which is characteristically higher for signal than for background processes; and the cosine of the angle between the missing momentum and the tag-side track in the $\mathrm{c.m.}$ frame, which, unlike background events, is expected to be positive for signal events due to the missing energy being only on the tag-side hemisphere. An additional kinematic variable we use is the energy asymmetry between the muon and the signal-side photon, $|E_{\mu}^{*} - E_{\gamma}^{*}| / (E_{\mu}^{*} + E_{\gamma}^{*})$, where the background is concentrated at high values while the signal is uniformly distributed in [0.0, 0.6], a range constrained by the photon energy threshold of $1$~GeV. We also use the total visible energy of the event, $E_{\mathrm{total}}^{*} = E_{\mu}^{*} + E_{\gamma}^{*} +E_{\mathrm{tag}}^{*}$, which separates the main backgrounds, as the signal clusters between 5.0 GeV and 10.0 GeV, while $e^{+}e^{-}\to\mu^{+}\mu^{-}$ and $e^{+}e^{-}\to\tau^{+}\tau^{-}$ events peak around 10.0 GeV and 6.0 GeV, respectively. 

Finally, as introduced in Ref.~\cite{Uno2021}, we use the $\xi$ variable, defined as
\begin{equation} \label{eq:xi_def}
    \xi=\frac{m^{2}_{\mathrm{miss}}-m_{\tau}^{2}-m_{\mathrm{track}}^2+\sqrt{s}~E_{\mathrm{track(tag)}}^{*}}{2\,|\vec{p}^{\hspace{0.6mm} *}_{\tau (\mathrm{tag})} | |\vec{p}^{\hspace{0.6mm} *}_{\mathrm{track (tag)}}|} \;,
\end{equation}
where $\vec{p}^{\hspace{0.6mm} *}_{\tau (\mathrm{tag})}$ and  $\vec{p}^{\hspace{0.6mm} *}_{\mathrm{track (tag)}}$ are the three-momenta of the tag $\tau$ and the tag-side track. The quantities $m_{\mathrm{track}}$, $E_{\mathrm{track(tag)}}^{*}$, and $P^{*}_{\mathrm{track(tag)}}=(E^{*}_{\mathrm{track(tag)}},\hspace{1mm} \vec{p}^{\hspace{0.6mm} *}_{\mathrm{track (tag)}})$ are the mass, the energy, and the four-momentum of the tag-side track, respectively. The quantity $m^{2}_{\mathrm{miss}}= [P^{*}_{\tau(\mathrm{tag})}-P^{*}_{\mathrm{track(tag)}}]^{2}$ is the missing mass squared, with $P^{*}_{\tau(\mathrm{tag})}=(E^{*}_{\tau(\mathrm{tag})}, \hspace{1mm}  \vec{p}^{\hspace{0.6mm} *}_{\tau (\mathrm{tag})})$ being the four-momentum of the tag $\tau$.  Since $\xi$ is computed from the reconstructed $m^{2}_{\mathrm{miss}}$, it can take values outside $[-1,1]$ for both signal and background, as shown in the lower panel of Figure~\ref{fig:xi_distribution}. The physical meaning of $\xi$ becomes apparent in the absence of missing energy, where Eq.~(\ref{eq:xi_def}) reduces to
\begin{equation} \label{eq:cos_tau_track}
    \xi=\frac{\vec{p}^{\hspace{0.6mm} *}_{\tau (\mathrm{tag)}} \cdot \vec{p}^{\hspace{0.6mm} *}_{\mathrm{track (tag)}}}{|\vec{p}^{\hspace{0.6mm} *}_{\tau (\mathrm{tag})} | |\vec{p}^{\hspace{0.6mm} *}_{\mathrm{track (tag})}|} =\cos{\theta_{\tau(\mathrm{tag}), \mathrm{track}(\mathrm{tag})}}\;,
\end{equation}
which is the cosine of the angle between the tag $\tau$ and the tag-side track in the $\mathrm{c.m.}$ frame. For fully reconstructed signal events, due to momentum conservation, the momentum of the tag $\tau$ can be written as $\vec{p}_{\tau (\mathrm{tag})}^{\hspace{0.6mm} *} = -\vec{p}_{\tau (\mathrm{sig})}^{\hspace{0.6mm} *}= -\vec{p}_{\mu}^{\hspace{0.6mm} *} -\vec{p}_{\gamma}^{\hspace{0.6mm}*} $, and therefore the signal events in the $\xi$ variable are expected to be distributed between [0, 1] and peak at 1. The  $\xi$  observable has the greatest discriminating power, exhibiting the highest feature importance score during the GBDT training process. It is followed, in descending order of importance, by the cosine of the angle between the missing momentum and the tag-side track; the missing momentum magnitude; the squared missing neutrino mass; the total visible energy; the tag-side charged particle $\mathrm{c.m.}$ momentum; the photon energy; the thrust; the energy asymmetry between the muon and the signal-side photon; and the number of neutral pions and photons in the tag-side hemisphere.

\begin{figure}[h!]
    \centering
    \includegraphics[width=0.98\linewidth]{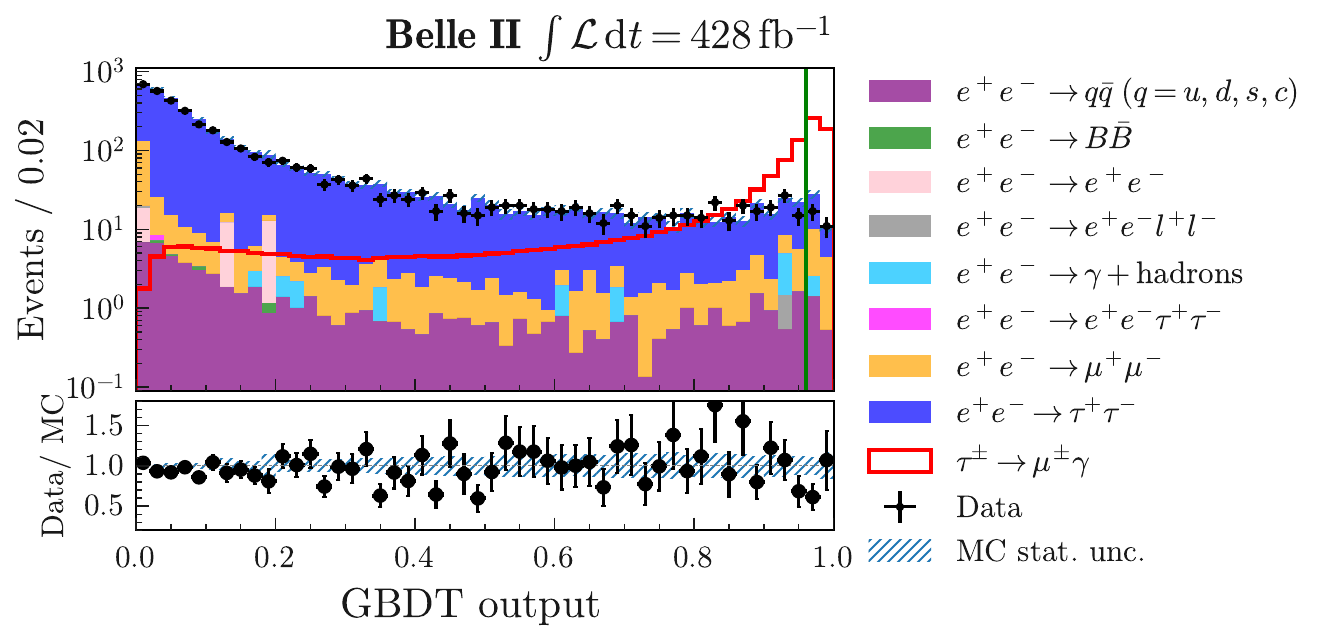}
    \caption{Distribution of the GBDT classifier output after applying the baseline selection (prior to the GBDT cut). Data and simulation include both the sideband and signal regions. Simulated background processes are stacked and normalized to an integrated luminosity of 428~fb$^{-1}$. A solid green line indicates the optimal selection threshold. The $\tau^{-} \to \mu^{-} \gamma$ signal simulation is overlaid with an arbitrary normalization factor for illustration. The lower panel displays the ratio of data to the total simulated background.}\label{fig:xgboost_distribution}
\end{figure}

All input variables used in the GBDT training are validated by comparing sideband data and simulation distributions. The ratio of sideband data to simulation is consistent with unity within uncertainties. The GBDT is trained and validated on independent simulation samples. No signs of overtraining or peak structures in the signal region vicinity $M_{\mathrm{bc}} \sim m_{\tau}$ and $ \Delta E / \sqrt{s} \sim 0$ are observed. In particular, all backgrounds remain uniformly distributed on $M_{\mathrm{bc}}$. The distribution of the GBDT classifier output is shown in Figure~\ref{fig:xgboost_distribution}. The optimal threshold on the GBDT output is 0.96, maximizing the figure of merit for a signal search at the 90\% C.L., $\epsilon_{\mathrm{sig}}/ (\frac{a}{2}+\sqrt{b})$, where $\epsilon_{\mathrm{sig}}$ and $b$ are the signal efficiency and the number of background events for a given selection, and $a=1.28$ represents the number of sigmas of a one-sided Gaussian function at the chosen significance~\cite{punzi2003}. The signal efficiency after the baseline and GBDT selections is $ 5.97\%$, while the background rejection reaches 99.8\% for $e^{+}e^{-}\to\tau^{+}\tau^{-}$, 97.9\% for $e^{+}e^{-}\to\mu^{+}\mu^{-}$, 98.7\% for $e^{+}e^{-}\to q\bar{q}$, and more than 99.9\% for other processes.

\section{Signal extraction model}\label{section4}

We perform a two-dimensional unbinned extended maximum-likelihood fit to the $M_{\mathrm{bc}}$ and $\Delta E / \sqrt{s}$ observables.
The extended likelihood is constructed as
\begin{equation}\label{eq:model_with_nuisance}
    L = \frac{e^{-(s+b)}}{N!} \prod^N_{i=1}(sS_i + bB_i) \prod_{k=1}^{K} \mathcal{G}({\alpha_{k}}; \mu_{k}, \sigma_{k}) \;,
\end{equation}
where $N=s+b$ is the number of observed events, and $s$ and $b$ are the numbers of signal and background events, respectively. The functions $S_{i} = S(M_{\mathrm{bc}}, \Delta E / \sqrt{s})_{i}$ and $B_{i} = B(M_{\mathrm{bc}}, \Delta E / \sqrt{s})_{i}$ are the probability density functions (PDFs) for signal and background, respectively, and the index $i$ runs over all $N$ events. The $\alpha_{k}$ represent nuisance parameters with Gaussian constraint terms  $\mathcal{G}(\alpha_{k};\mu_{k},\sigma_{k})$. The index $k$ runs over all $K$ nuisance parameters, with $\mu_{k}$ and $\sigma_{k}$ denoting the nominal value and total systematic uncertainty of the $k$-th parameter, respectively. Section~\ref{section5} provides further details on the nuisance parameters and their constraints.

The signal yield is expressed as a function of  the branching fraction $\mathcal{B}(\tau^{-}\to\mu^{-}\gamma)$, which is the parameter of interest, using the relation
\begin{equation}\label{eq:br}
    s= \mathcal{B}(\tau^{-}\to\mu^{-}\gamma) \cdot 2\cdot  \mathcal{L}_{\mathrm{int}}\cdot \epsilon_{\mathrm{sig}} \cdot \sigma(e^{+}e^{-}\to \tau^{+} \tau^{-})\;,
\end{equation}
where $\epsilon_{\mathrm{sig}}$ is the signal efficiency, $\mathcal{L}_{\mathrm{int}}$ is the total integrated luminosity, and $\sigma(e^{+}e^{-}\to\tau^{+} \tau^{-})$ is the $\tau^{+} \tau^{-}$ production cross-section.

The signal and background PDFs are constructed from the simulated samples as the products of the one-dimensional models of $M_{\mathrm{bc}}$ and $\Delta E / \sqrt{s}$, since the correlation between these variables is small, with Pearson correlation coefficients below 0.1 (0.2) for the background (signal) sample. The signal PDF is constructed as $S(M_{\mathrm{bc}}$, $\Delta E / \sqrt{s}$) $= S^{'}(M_{\mathrm{bc}})$ $\times$ $S^{''}(\Delta E / \sqrt{s})$, where each model $S^{'}$ and $S^{''}$ is the sum of two bifurcated Gaussian distributions with a common mean. The parameter values and uncertainties are presented in Table  \ref{table:signal_parameters}.
The average resolutions of $\Delta E / \sqrt{s}$ and  $M_{\mathrm{bc}}$ are obtained from the left and right effective standard deviations, defined as the square root of the average variances. The resulting values are $\sigma_{\Delta E / \sqrt{s}} = (3.7 \pm 0.1)\times 10^{-3}$ and $\sigma_{M_{\mathrm{bc}}} = 8.3 \pm 0.3~\mathrm{MeV}/c^{2}$. The signal region as introduced at the end of Section~\ref{section3} corresponds to a $\pm 5\sigma_{M_{\mathrm{bc}}}$ window around the $\tau$ mass for the $M_{\mathrm{bc}}$ observable and $\pm 5.5\sigma_{\Delta E / \sqrt{s}}$ window around the mean for the $\Delta E / \sqrt{s}$ observable. Similarly to the signal model, the background PDF is defined as $B(M_{\mathrm{bc}}, \Delta E / \sqrt{s}) = B^{'}(M_{\mathrm{bc}}) \times B^{''}(\Delta E / \sqrt{s})$, where $B^{'}$ is modeled as a uniform distribution, as no peaking structure is foreseen from simulation, and $B^{''}$ is modeled as the sum of the two main background components $e^{+}e^{-}\to\tau^+\tau^-$ and $e^{+}e^{-}\to\mu^{+}\mu^{-}$, since all other backgrounds are expected to be negligible from simulation. In the $\Delta E / \sqrt{s}$ observable, the background model is 
\begin{equation}
    B^{''}(\Delta E / \sqrt{s}) = f J_{\tau^{+}\tau^{-}}(\Delta E / \sqrt{s}) + (1-f) J_{\mu^{+}\mu^{-}}(\Delta E / \sqrt{s})\;,
\end{equation}
 where \textit{f} is the fraction of the $\tau^+\tau^-$ component and the individual PDFs $J_{\tau^{+}\tau^{-}}$ and  $J_{\mu^{+}\mu^{-}}$ are modeled by Johnson's $S_{U}$ distributions \cite{johnsonsu}. The central values and uncertainties of the parameters of both models are presented in Table \ref{table:background_parameters}. The parameters of the signal PDF are fixed to the values obtained from simulation (see Table \ref{table:signal_parameters}). For the background PDFs, the central values of the parameters listed in Table \ref{table:background_parameters} are used, and the uncertainties are taken into account as described in Section~\ref{section5}. In the fit to data, the branching fraction, the background yield $b$, and the fraction $f$ are left free. The branching fraction is allowed to take negative values.

 Before unblinding the signal region, the expected $N_{\tau^{+}\tau^{-}}$ and $N_{\mu^{+}\mu^{-}}$ yields are extrapolated by integrating the PDFs with the $b$ and $f$ parameters determined from the fit to sideband data. The sidebands contain 22 observed events, consistent with the simulation expectation of $28.0 \pm 2.8$ events. The fit to the sideband data yields an expected background of $5.7 \pm 1.2$ events in the signal region, consistent with the estimate of $7.3 \pm 0.7$ events obtained from the same procedure applied to the simulated sample. The expected $\tau^+\tau^-$ background fraction is $f=0.51 \pm 0.17$ from the fit to the sideband data and $f=0.71 \pm 0.07$ obtained from the same procedure applied to the simulated sample. By directly counting the simulated events in the signal region, $N_{\tau^{+}\tau^{-}}= 4.5 \pm 1.1$ events are found, of which the two dominant $\tau$ background processes are $\tau^{-} \to \mu^{-} \bar{\nu}_{\mu} \nu_{\tau}$ and $\tau^{-} \to \pi^{-} \nu_{\tau}$ together with a photon from ISR, contributing 81\% and 19\%, respectively.

\begin{table}[h]
\centering
\caption{Signal model parameters of $S^{'}(M_{\mathrm{bc}})$ and $S^{''}(\Delta E / \sqrt{s})$ obtained from a fit to simulated signal samples. Each distribution is fitted with the sum of two bifurcated Gaussian functions sharing a common mean. The fraction corresponds to the first Gaussian component, which has widths $\sigma_{1}^{\text{left}}$ and $\sigma_{1}^{\text{right}}$; the second Gaussian component has widths $\sigma_{2}^{\text{left}}$ and $\sigma_{2}^{\text{right}}$.}
\label{table:signal_parameters}
\vspace{1mm}
\scriptsize{
\begin{tabular}{p{4cm} p{4cm} p{4cm}}
\hline
Parameter & $S^{'}(M_{\mathrm{bc}})$  & $S^{''}(\Delta E / \sqrt{s})$\\
\hline
\hline
fraction                       & $0.58 \pm 0.03$ & \hspace{1.2mm}$0.73 \pm 0.04$ \\
mean                      & \hspace{-4.5mm}$1777.46 \pm 0.07$ MeV$/c^{2}$ & \hspace{-1.5mm}$(0.17 \pm 0.04)\times 10^{-3}$\\
$\sigma^{\mathrm{left}}_{1}$       & $9.4 \pm 0.2$ MeV$/c^{2}$ & $(2.8 \pm 0.2)\times10^{-3}$ \\
$\sigma^{\mathrm{left}}_{2}$             & $5.9 \pm 0.1$ MeV$/c^{2}$ & $(4.46 \pm 0.05)\times10^{-3}$\\
$\sigma^{\mathrm{right}}_{1}$                & $11.8 \pm 0.4$ MeV$/c^{2}$ & $(4.5 \pm 0.1)\times10^{-3}$ \\
$\sigma^{\mathrm{right}}_{2}$           & $6.0 \pm 0.1$ MeV$/c^{2}$ & $(2.77 \pm 0.05)\times10^{-3}$ \\

\hline
\end{tabular}
}
\end{table}

\begin{table}[h]
\centering
\caption{Background model parameters for the $\Delta E / \sqrt{s}$ observable obtained from fits to simulated $\tau^{+}\tau^{-}$ and $\mu^{+}\mu^{-}$ samples. Each background component is described by a Johnson $S_U$ distribution with location parameter $\mu$, scale parameter $\lambda$ ($\lambda > 0$), and shape parameters $\gamma$ and $\delta$ ($\delta > 0$).}
\vspace{1mm}
\label{table:background_parameters}
\scriptsize{
\begin{tabular}{p{4cm} p{4cm} p{4cm}}
\hline
Parameter & $J_{\mu^{+}\mu^{-}}$ & $J_{\tau^{+}\tau^{-}}$ \\
\hline
\hline
$\delta$ & \hspace{0.6mm}$1.3 \pm 0.8$  & $0.8 \pm 0.2$ \\
$\gamma$ & \hspace{0.6mm}$1.1 \pm 1.0$  & \hspace{-2.3mm}$-0.5 \pm 0.2$\\
$\mu$    & \hspace{-0.5mm}$(1.5 \pm 0.4)\times10^{-2}$  & \hspace{-3.4mm}$(-5.9 \pm 1.2)\times10^{-3}$ \\
$\lambda$& \hspace{-0.5mm}$(4.4 \pm 3.3)\times10^{-3}$  &\hspace{-1.1mm}$(3.6 \pm 0.8)\times10^{-3}$ \\
\hline
\end{tabular}
}
\end{table}

The two-dimensional fit is performed in an extended region defined by $M_{\mathrm{bc}} \in [1.7, 1.85]$ GeV/$c^{2}$ and $\Delta E / \sqrt{s} \in [-0.02, 0.02]$, referred to as the fit region. 
Within the fit region, the background distribution in $M_{\mathrm{bc}}$ is well described by the uniform model, as confirmed by large-sample simulation studies, ensuring a stable determination of the background shapes. The signal efficiency in the fit region is $\epsilon_{\mathrm{sig}} = 5.96 \pm 0.01 (\mathrm{stat}) \%$. In the fit region, the expected background yield is $15.7 \pm 3.4$, obtained by summing the 10 observed events in the sidebands within the fit region and the signal-region expected background. The corresponding background yields are $N_{\tau^{+}\tau^{-}}= 8.0 \pm 3.6$ and $N_{\mu^{+}\mu^{-}} = 7.7 \pm 4.6$; these values are used in the analysis to calculate a pre-fit expected upper limit. Applying the same procedure to the simulated sample, the expected background is $13.1 \pm 1.4$, with $N_{\tau^{+}\tau^{-}} = 9.3 \pm 1.6$ and $N_{\mu^{+}\mu^{-}}=3.8 \pm 1.4$, consistent with the data-based estimate within uncertainties. A summary of all the expected event yields discussed above, together with the observed events, is given in Table~\ref{table:number_events}. We obtain the pre-fit expected upper limit $\mathcal{B}(\tau^{-}\to\mu^{-}\gamma)^{\mathrm{exp}}_{\mathrm{pre\text{-}fit}} < 5.8$ $(8.7)\times 10^{-8}$ at the 90\% (95\%)~C.L., including the systematic uncertainties discussed in Section~\ref{section5} and the CL$_{s}$ technique presented in Section~\ref{section6}.
\section{Systematic Uncertainties}\label{section5}

Systematic uncertainties arise from several sources, such as uncertainties from external measurements or theoretical inputs, and differences between experimental data and simulation due to possible mismodeling in the generation and reconstruction of the simulated samples. We consider systematic uncertainties affecting the total integrated luminosity, the $\tau^{+}\tau^{-}$ production cross-section, the signal efficiency, and the background modeling. Table~\ref{table:summary_systematics} summarizes all systematic uncertainties.

\begin{table}[h!]
\centering
\caption{Systematic uncertainties considered in the analysis. For each uncertainty source, the affected parameter and the associated relative uncertainty [\%] are provided.}

\begin{tabular}{p{3.8cm} p{4cm} p{6.5cm}}
\hline
Uncertainty source & Affected parameter & Relative uncertainty [\%] \\
\hline 
\hline
Simulated samples size  &$\epsilon_{\mathrm{sig}}$&  0.2\% \\
Trigger efficiency &$\epsilon_{\mathrm{sig}}$&  1.0\% \\
Lepton identification  &$\epsilon_{\mathrm{sig}}$&  0.6\% \\
Photon efficiency  &$\epsilon_{\mathrm{sig}}$&   0.8\% \\
Tracking efficiency  &$\epsilon_{\mathrm{sig}}$&  0.5\% \\
GBDT  &$\epsilon_{\mathrm{sig}}$&  5.7\% \\
$L_{\mathrm{int}}$ &-& 0.5\% \\
$\sigma(e^{+} e^{-} \to \tau^{+} \tau^{-})$ &-&  0.7\% \\
$J_{\mu^{+}\mu^{-}}$  model & $\delta$, $\gamma$, $\mu$, $\lambda$ & 55\%, 89\%, 23\%, 49\%\\
$J_{\tau^{+}\tau^{-}}$ model &  $\delta$, $\gamma$, $\mu$, $\lambda$  & 16\%, 48\%, 20\%, 16\%\\
\hline
\end{tabular}
\label{table:summary_systematics}
\end{table}

The total systematic uncertainty on the product of the integrated luminosity, the $\tau^{+}\tau^{-}$ production cross-section, and the signal efficiency is implemented as a nuisance parameter in the likelihood model, assumed to follow a Gaussian function with mean given by $2 \cdot  \mathcal{L}_{\mathrm{int}}\cdot \epsilon_{\mathrm{sig}} \cdot \sigma(e^{+}e^{-}\to \tau^{+} \tau^{-})$ and a corresponding width of 6.0\%, which corresponds to the quadratic sum of the individual sources.

An uncertainty of 0.5\% on the integrated luminosity is considered. The value $\mathcal{L}_{\mathrm{int}}=428\pm2$~fb$^{-1}$ is obtained in Ref.~\cite{luminosity_belle2} using three independent methods based on $e^{+}e^{-} \to e^{+}e^{-}(n\gamma)$, $e^{+}e^{-} \to \gamma\gamma(n\gamma)$, and $e^{+}e^{-} \to \mu^{+}\mu^{-}(n\gamma)$ processes, where $n$ represents inclusive photon multiplicity. 

For the $\tau^{+}\tau^{-}$ production cross section, we use the precisely calculated value  $\sigma = 0.919 \pm 0.003~\text{nb}$ \cite{cross_tau} at the c.m. energy of $ \sqrt {s} = 10.58$ GeV, as the basis for extrapolation to the $\mathrm{c.m.}$ energies relevant to this analysis. The resulting weighted-average cross section is $0.92~\text{nb}$, with an assigned systematic uncertainty of 0.7\%.

Possible discrepancies between simulated samples and data contribute as uncertainty sources in the signal efficiency. The sources of uncertainty in signal efficiency are the finite size of simulated samples, the trigger efficiency correction, the lepton identification efficiency, the tracking efficiency, the photon-detection efficiency, and the GBDT efficiency. The systematic uncertainty on the ECL-based trigger efficiency is evaluated by comparing sideband data and simulation, with orthogonal CDC-based triggers serving as the reference. We assign the difference as the systematic uncertainty. The systematic uncertainty associated with the corrections to the simulated muon-identification efficiency is derived from auxiliary measurements in data using $J/\psi \to \mu^{+} \mu^{-}$, $e^{+}e^{-} \to \mu^{+} \mu^{-}$, and $e^{+}e^{-} \to e^{+}e^{-} \mu^{+} \mu^{-}$ \cite{lidmilesi}. These corrections are obtained as functions of momentum, polar angle, and charge of the muon candidate, and are applied to simulated events. We evaluate the systematic uncertainty by varying the corrections within their uncertainties and calculating the resulting change in selection efficiency. The track-reconstruction efficiency is measured in data and simulation using $e^{+}e^{-} \to \tau^{+} \tau^{-}$. A systematic uncertainty of 0.27\% is assigned per track, resulting in a total uncertainty of 0.5\% for both the signal and tag charge particles. The systematic uncertainty on the photon detection efficiency, measured with $e^{+}e^{-} \to \mu^{+} \mu^{-}\gamma$ events, is calculated from the data-simulation efficiency ratio. We derive a systematic uncertainty for the GBDT selection by applying the classifier to a control sample of $\tau^{-} \to \pi^{-} \nu_{\tau}$ decays accompanied by a radiated photon in the same hemisphere. In this control sample, we remove the muon identification ($P_{\mu} > 0.95$) requirement for the signal muon and instead require a pion identification ($P_{\pi} > 0.9$); all other selection criteria are maintained. The selection criteria on the GBDT output are tuned so that the GBDT selection efficiency in the control sample matches that of the $\tau^{-} \to \mu^{-}\gamma$ signal process. We then compute the discrepancy in the efficiency between data and MC as the difference of the ratio $\epsilon^{\mathrm{GBDT},\pi\gamma}_{\mathrm{data}}/\epsilon^{\mathrm{GDBT}, \pi\gamma}_{\mathrm{MC}}$ from unity. The observed deviation  $1-({\epsilon^{\mathrm{GBDT},\pi\gamma}_{\mathrm{data}}} /{\epsilon^{\mathrm{GBDT},\pi\gamma}_{\mathrm{MC}}}) = 5.7\%$ between data and simulation in the control sample is then assigned as the systematic uncertainty. 
Since the effect of ISR is independent of the final state, from the same control sample, the potential mismodeling of ISR in the different $\Delta E / \sqrt{s} $ resolution between data and MC can be checked. No significant difference is observed between the data and MC distributions in this control sample; therefore, no systematic contribution is assigned.
The effect of FSR in $\tau$ decays, simulated by PHOTOS, depends on the final-state particles and could affect signal modeling. 
Since the final state contains a single muon, the impact of final-state radiation on the signal PDF is negligible. This was confirmed by comparing the signal distributions with and without bremsstrahlung photon recovery, yielding a KS distance below 0.1\%. Therefore, no additional systematic uncertainty is assigned.

The systematic uncertainty associated with the background modeling is evaluated by constructing two alternative background models via simultaneous shifts of the background PDF parameters and incorporating them into the likelihood model with a shape-related nuisance parameter. The parameters and their uncertainties are given in Table~\ref{table:background_parameters}. To account for correlations among the parameters, the variation is performed using an eigen‑decomposition of the covariance matrix. We use the principal eigenvector, corresponding to the dominant source of uncertainty, to apply simultaneous shifts to all parameters. Two alternative background models are constructed by varying one standard deviation ($\sigma$). The resulting percentage shifts for the ($\delta$, $\gamma$, $\mu$, $\lambda$) parameters are ($55\%$, $89\%$, $23\%$, $49\%$) for the $J_{\mu^{+}\mu^{-}}$ model and ($16\%$, $48\%$, $20\%$, $16\%$) for the $J_{\tau^{+}\tau^{-}}$ model. This uncertainty is then incorporated into the likelihood model by introducing a shape-related nuisance parameter with a Gaussian constraint, which allows smooth transitions between the $-1\sigma$, nominal, and $+1\sigma$ background models.

The systematic uncertainty associated with the limited statistical precision of the signal model is evaluated using the same procedure as for the background models. It is found to be negligible given the large available sample size.
The signal model shape can also be affected by detector effects. The photon energy scale, as well as possible biases in the charged-track momentum scale and energy losses due to imperfections in the magnetic-field description, are corrected in data to ensure agreement with the simulation. Other effects, such as differences in photon energy and angular resolutions between data and simulation, have been studied using radiative $e^{+}e^{-}\to\mu^{+}\mu^{-}\gamma$ events in Ref.~\cite{alppaper}, which reports a 17\% larger energy resolution in data and a 9\% larger angular resolution in simulation. To assess the impact of these resolution differences, alternative signal models are constructed and incorporated into the likelihood through an additional shape-related nuisance parameter, $\alpha_{\mathrm{sig}}$, with a Gaussian constraint centered at zero with a width of one, following the procedure used for the background models. The fit yields $\alpha_{\mathrm{sig}}=0.1\pm 1.0$, consistent with the nominal signal model. The introduction of this additional parameter has a negligible impact on the fitted branching fraction and the other fit parameters. The observed upper limit remains unchanged at the quoted precision, and the expected upper limits are consistent with the nominal results. Therefore, no additional systematic uncertainty is assigned for these effects.

The systematic uncertainties are incorporated into the likelihood model via nuisance parameters, which are profiled in the construction of the upper limit during the calculation of the likelihood ratio test statistic.

\section{Result}\label{section6}
\begin{figure}[h]
    \centering
    \includegraphics[width=0.9\linewidth]{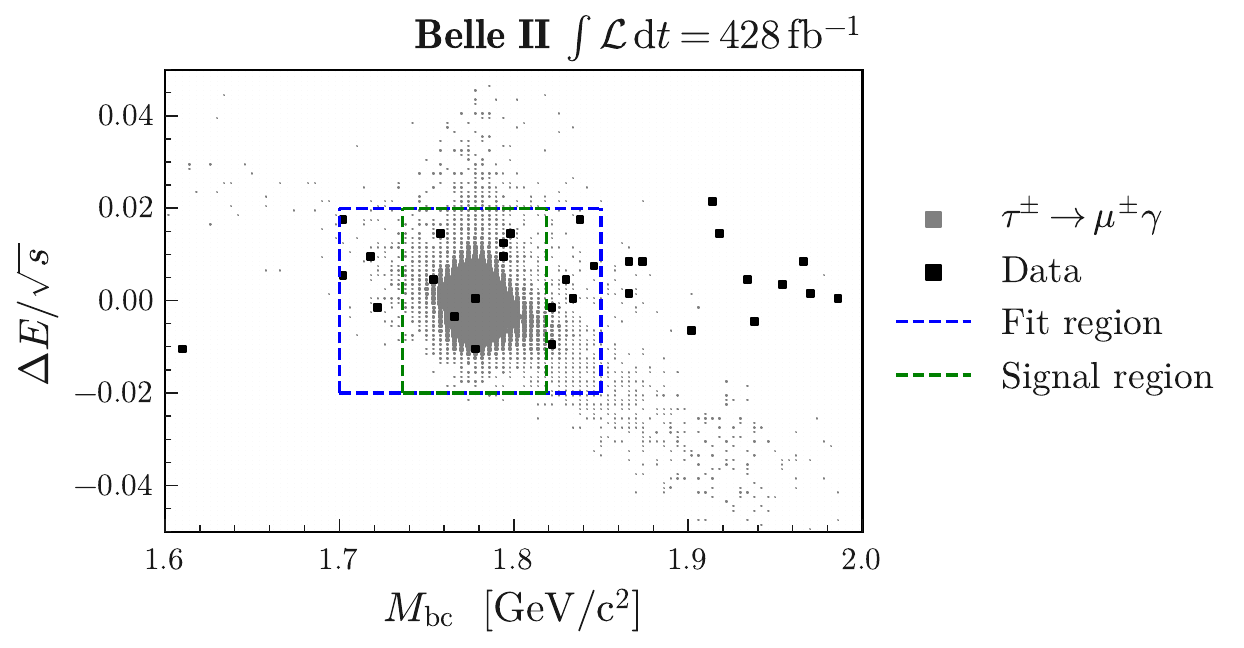}
    \caption{Distribution of data (black) and simulated $\tau^{-} \to \mu^{-}\gamma$ signal events (gray) in the two-dimensional plane of $M_{\mathrm{bc}}$ versus $\Delta E / \sqrt{s}$. The green dashed lines mark the boundaries of the signal region, which is kept hidden until the final stage of the analysis. The blue dashed rectangle denotes the fit region used in the likelihood fit to extract the final results.}
    \label{fig:mc_data_signalregion}
\end{figure}

After unblinding the data, 8 events are observed in the signal region and 18 events in the fit region. Figure~\ref{fig:mc_data_signalregion} shows the data distribution together with the simulated signal, with the fit and signal regions indicated. Table~\ref{table:number_events} summarizes the observed events together with the expected background yields in the signal, fit, and sideband regions, as described in Section~\ref{section4}.

The determination of the  $\tau^{-}\to\mu^{-}\gamma$ yield is performed using a two-dimensional unbinned extended maximum-likelihood fit. 
No significant excess over the background prediction from the SM is observed, and the resulting branching fraction $\mathcal{B}(\tau^{-}\to\mu^{-}\gamma) =(2.8 ^{+4.4}_{-2.6}) \times 10^{-8}$ is consistent with zero, where the uncertainty includes both the statistical and systematic contributions. The result corresponds to  $\tilde{s}=1.3^{+2.0}_{-1.3}$ signal, and $\tilde{b}=16.7 ^{+ 4.6}_{-3.9}$ background events in the fit region, consistent with the 18 observed events.
The fitted $\tau^{+}\tau^{-}$ background fraction is $f=0.34 \pm 0.17$, corresponding to $N_{\tau^{+}\tau^{-}}=6.1 \pm 3.8$ and $N_{\mu^{+}\mu^{-}}=11.9 \pm 4.6$ in the fit region. The sensitivity is then recalculated using the post-fit values derived from the unblinded data as presented below. The fit projections are shown in Figure~\ref{fig:cls_fitsidebands}.

\begin{table}[htpb]
\centering
\caption{Observed and expected number of events in the different $M_{\mathrm{bc}}$ regions. The expectations are obtained from sideband extrapolation performed on data and from the same procedure applied to the simulated sample. The observable $\Delta E/\sqrt{s}$ $\in$ $[-0.02,0.02]$ in all regions.
}\label{table:number_events}
\vspace{1mm}
\scriptsize{
\begin{tabular}{p{4cm} p{2.6cm} p{2.4cm} p{4.2cm}}
\hline
  &  Signal region   &  Fit region   & Sideband region \\
 & $M_{\mathrm{bc}}$ $\in$ [1.736, 1.819]   & $M_{\mathrm{bc}}$ $\in$  [1.7, 1.85]   &  $M_{\mathrm{bc}}$ $\in$ [1.6, 1.736] $\cup$ [1.819, 2.0] \\
\hline
\hline
Data& $8.0 \pm 2.8$ & $18.0 \pm 4.2$ & $22.0 \pm 4.7$ \\
\scriptsize{Expected~from~sideband~data}& $5.7 \pm 1.2$ & $15.7 \pm 3.4$ & $22.0 \pm 4.7$ \\
\scriptsize{Expected~from~simulation}& $7.3 \pm 0.7$ & $13.1 \pm 1.4 $& $28.0 \pm 2.8$\\

\hline

\end{tabular}
}

\end{table}

We construct an upper limit on $\mathcal{B}(\tau^- \to \mu^- \gamma)$ using a modified frequentist approach based on the CL${_s}$ technique \cite{cls_method, Junk_1999}. The CL${_s}$ method is implemented using the RooStats framework and the FrequentistCalculator class \cite{moneta2011roostatsproject, Brun:1997pa}. The profile-likelihood ratio is adopted as the test statistic. For each of 20 tested values of the branching fraction in the range $[0,\, 20 \times 10^{-8}]$, we generate $10^4$ pseudo-experiments under both the signal-plus-background and background-only hypotheses. The background-only generation model, including the nominal values of its nuisance parameters, remains unchanged across all tested branching fractions. The signal-plus-background generation model is updated according to the branching fraction being tested. In generating these pseudo-experiments, the nuisance parameters are set to their nominal values, while their uncertainties, as determined in Section \ref{section5}, are incorporated into the subsequent likelihood fits. Each pseudo-experiment is fitted twice: once with the parameter of interest (the branching fraction) fixed to the tested value, and once where it is allowed to float. In both cases, the nuisance parameters are profiled in the fit. Validation of the fit procedure using pull distributions and profile likelihood plots shows no evidence of bias. At each tested branching fraction, the profile-likelihood test statistic is evaluated for the observed data and for pseudo-experiments generated under both the signal-plus-background and background-only hypotheses. The observed CL$_s$ value is then calculated by comparing the observed test statistic with the two corresponding toy distributions. The observed upper limit is obtained from the branching fraction at which the CL$_s$ value crosses the threshold corresponding to the confidence level. The expected upper limits are determined from the background-only pseudo-experiments by applying the same procedure to each pseudo-experiment. The median of the resulting distribution of upper limits is taken as the expected upper limit, while its central intervals characterize the expected variation. The thresholds are CL$_s=0.10$ and CL$_s=0.05$ for the 90\% and 95\% confidence levels, respectively.

\begin{figure}[h]
    \centering
    \includegraphics[width=.99\linewidth]{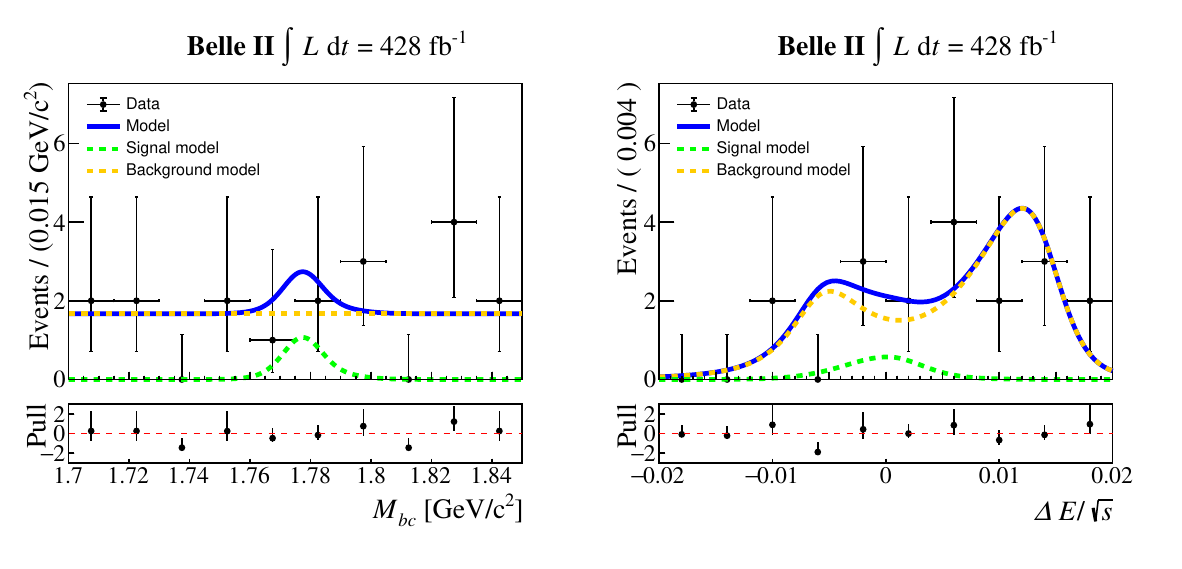}
    \caption{Projections of the fit to data. The left and right panels show the $M_{\mathrm{bc}}$ and $\Delta E / \sqrt{s}$ observables, respectively, with the individual signal and background contributions from the fit overlaid. The pulls, defined as the residual divided by the data uncertainty per bin, are shown in the bottom panels.}
    \label{fig:cls_fitsidebands}
\end{figure}

After unblinding, we evaluate the expected upper limit using the profiled values of $b$, $f$, and the other nuisance parameters obtained from the fit to the observed data. Pseudo-experiments are then generated under the background-only hypothesis using these parameter values. We determine the expected upper limit from the median of the resulting CL$_s$ distribution. The expected upper limit on the branching fraction at the 90\% (95\%)~C.L. is:
\begin{equation}\label{eq:expected_ul_cls}
\mathcal{B}(\tau^{-}\to\mu^{-}\gamma)^{\mathrm{exp}}_{\mathrm{post\text{-}fit}} < 7.6 \hspace{1mm}(9.2)\times 10^{-8}\,.
\end{equation}

The observed upper limit on the branching ratio at the 90\% (95\%) C.L. is:
\begin{equation}\label{eq:observed_ul_cls}
    \mathcal{B}(\tau^{-}\to\mu^{-}\gamma)^{\mathrm{obs}} < 9.5 \hspace{1mm}(12.2)\times10^{-8}\,.
\end{equation} 

The observed result lies within the expected CL$_{s}$ bands, with the deviation from the central value attributable to statistical fluctuation. Compared with the previous Belle search, this analysis achieves a 41\% higher signal efficiency and an 81\% reduction in the expected background yield within the signal region defined in Ref.~\cite{Uno2021}.

\section{Summary}\label{section7}

We report the search for the $\tau^{-}\to\mu^{-}\gamma$ decay at the Belle II experiment, using a data sample corresponding to an integrated luminosity of 428 fb$^{-1}$. The analysis strategy employs a GBDT-based selection to suppress background contributions. For the first time in the $\tau^{-}\to\mu^{-}\gamma$ search, we construct a unified likelihood function that encapsulates the entire analysis, treating systematic uncertainties as nuisance parameters. An unbinned extended maximum-likelihood fit is performed in the two-dimensional space of $M_{\mathrm{bc}} $ and $ \Delta E/\sqrt{s}$. The measured branching fraction, $\mathcal{B}(\tau^{-}\to\mu^{-}\gamma) =(2.8 ^{+4.4}_{-2.6}) \times 10^{-8}$ is consistent with zero. Upper limits on $\mathcal{B}(\tau^{-} \to \mu^{-} \gamma)$ are derived using the CL${_s}$ technique, which provides a conservative limit in the case of a 
downward fluctuation of the background yield. The expected upper limits are $\mathcal{B}(\tau^{-}\to\mu^{-}\gamma) < 7.6$ $(9.2)\times10^{-8}$ at the 90\% (95\%) C.L., while the observed upper limits are found to be $\mathcal{B}(\tau^{-}\to\mu^{-}\gamma) < 9.5$ $(12.2)\times10^{-8}$ at 90\% (95\%) C.L. 

With improved signal selection, enhanced background suppression, and a likelihood-based statistical framework that naturally incorporates systematic uncertainties, this analysis establishes a foundation for future $\tau^{\pm} \to \ell^{\pm} \gamma$ searches at Belle II with larger data samples.

\section*{Acknowledgements}
\sloppy
This work, based on data collected using the Belle II detector, which was built and commissioned prior to March 2019,
was supported by
Higher Education and Science Committee of the Republic of Armenia Grant No.~23LCG-1C011;
Australian Research Council and Research Grants
No.~DP200101792, 
No.~DP210101900, 
No.~DP210102831, 
No.~DE220100462, 
No.~LE210100098, 
and
No.~LE230100085; 
Austrian Federal Ministry of Education, Science and Research,
Austrian Science Fund (FWF) Grants
DOI:~10.55776/P34529,
DOI:~10.55776/J4731,
DOI:~10.55776/J4625,
DOI:~10.55776/M3153,
and
DOI:~10.55776/PAT1836324,
and
Horizon 2020 ERC Starting Grant No.~947006 ``InterLeptons'';
Natural Sciences and Engineering Research Council of Canada, Digital Research Alliance of Canada, and Canada Foundation for Innovation;
National Key R\&D Program of China under Contract No.~2024YFA1610503,
and
No.~2024YFA1610504
National Natural Science Foundation of China and Research Grants
No.~11575017,
No.~11761141009,
No.~11705209,
No.~11975076,
No.~12135005,
No.~12150004,
No.~12161141008,
No.~12405099,
No.~12475093,
and
No.~12175041,
and Shandong Provincial Natural Science Foundation Project~ZR2022JQ02;
the Czech Science Foundation Grant No. 22-18469S,  Regional funds of EU/MEYS: OPJAK
FORTE CZ.02.01.01/00/22\_008/0004632 
and
Charles University Grant Agency project No. 246122;
European Research Council, Seventh Framework PIEF-GA-2013-622527,
Horizon 2020 ERC-Advanced Grants No.~267104 and No.~884719,
Horizon 2020 ERC-Consolidator Grant No.~819127,
Horizon 2020 Marie Sklodowska-Curie Grant Agreement No.~700525 ``NIOBE''
and
No.~101026516,
and
Horizon Europe Marie Sklodowska-Curie Staff Exchange project JENNIFER3 Grant Agreement No.~101183137 (European grants);
L’Institut National de Physique Nucl\'eaire et de Physique des
Particules (IN2P3) du CNRS under Project Identification No.
CNRS-IN2P3-14-PP-033
and L’Agence Nationale de la Recherche (ANR) under Grant No. ANR-23-CE31-
0018 and ANR-25-CE31-1333 (France);
BMFTR, DFG, HGF, MPG, and AvH Foundation (Germany);
Department of Atomic Energy under Project Identification No.~RTI 4002,
Department of Science and Technology,
and
UPES SEED funding programs
No.~UPES/R\&D-SEED-INFRA/17052023/01 and
No.~UPES/R\&D-SOE/20062022/06 (India);
Israel Science Foundation Grant No.~2476/17,
U.S.-Israel Binational Science Foundation Grant No.~2016113, and
Israel Ministry of Science Grant No.~3-16543;
Istituto Nazionale di Fisica Nucleare and the Research Grants BELLE2,
and
the ICSC – Centro Nazionale di Ricerca in High Performance Computing, Big Data and Quantum Computing, funded by European Union – NextGenerationEU;
Japan Society for the Promotion of Science, Grant-in-Aid for Scientific Research Grants
No.~16H03993,
No.~16H06492,
No.~16K05323,
No.~17H01133,
No.~17H05405,
No.~18K03621,
No.~18H03710,
No.~18H05226,
No.~19H00682, 
No.~20H05850,
No.~20H05858,
No.~22H00144,
No.~22K14056,
No.~22K21347,
No.~23H05433,
No.~26220706,
No.~26400255,
and
No.~26H02056,
and
the Ministry of Education, Culture, Sports, Science, and Technology (MEXT) of Japan;  
National Research Foundation (NRF) of Korea Grants
No.~2021R1-F1A-1064008,
No.~2022R1-A2C-1003993,
No.~RS-2018-NR031074,
No.~RS-2021-NR060129,
No.~RS-2024-00354342,
No.~RS-2025-02219521,
No.~RS-2026-25471491,
No.~RS-2026-25480677,
and
No.~RS-2026-25486791,
Radiation Science Research Institute,
Foreign Large-Size Research Facility Application Supporting project,
the Global Science Experimental Data Hub Center, the Korea Institute of Science and
Technology Information (K26L1M2C3)
and
KREONET/GLORIAD;
Universiti Malaya RU grant, Akademi Sains Malaysia, and Ministry of Education Malaysia;
Frontiers of Science Program Contracts
No.~FOINS-296,
No.~CB-221329,
No.~CB-236394,
No.~CB-254409,
and
No.~CB-180023, and SEP-CINVESTAV Research Grant No.~237 (Mexico);
the Polish Ministry of Science and Higher Education and the National Science Center;
the Ministry of Science and Higher Education of the Russian Federation
and
the HSE University Basic Research Program, Moscow;
University of Tabuk Research Grants
No.~S-0256-1438 and No.~S-0280-1439 (Saudi Arabia);
Slovenian Research Agency and Research Grants
No.~J1-50010
and
No.~P1-0135;
Ikerbasque, Basque Foundation for Science,
State Agency for Research of the Spanish Ministry of Science and Innovation through Grant No. PID2022-136510NB-C33, Spain,
the Severo Ochoa project CEX2023-001292-S funded by MICIU/AEI, State Secretariat for
Telecommunications and Digital Infrastructure with reference
TSI-069100-2023-0012, State Agency for Research of the Spanish Ministry
of Science, Innovation and Universities through Grant No
PID2024-156645NB-C21;
The Knut and Alice Wallenberg Foundation (Sweden), Contracts No.~2021.0174, No.~2021.0299, and No.~2023.0315;
National Science and Technology Council,
and
Ministry of Education (Taiwan);
Thailand Center of Excellence in Physics;
TUBITAK ULAKBIM (Turkey);
National Research Foundation of Ukraine, Project No.~2020.02/0257,
and
Ministry of Education and Science of Ukraine;
the U.S. National Science Foundation and Research Grants
No.~PHY-1913789 
and
No.~PHY-2111604, 
and the U.S. Department of Energy and Research Awards
No.~DE-AC06-76RLO1830, 
No.~DE-SC0007983, 
No.~DE-SC0009824, 
No.~DE-SC0009973, 
No.~DE-SC0010007, 
No.~DE-SC0010073, 
No.~DE-SC0010118, 
No.~DE-SC0010504, 
No.~DE-SC0011784, 
No.~DE-SC0012704, 
No.~DE-SC0019230, 
No.~DE-SC0021616, 
No.~DE-SC0022350, 
No.~DE-SC0023470; 
and
the Vietnam Academy of Science and Technology (VAST) under Grant
No.~DL0000.05/26-27.

These acknowledgements are not to be interpreted as an endorsement of any statement made
by any of our institutes, funding agencies, governments, or their representatives.

We thank the SuperKEKB team for delivering high-luminosity collisions;
the KEK cryogenics group for the efficient operation of the detector solenoid magnet and IBBelle on site;
the KEK Computing Research Center for on-site computing support; the NII for SINET6 network support;
and the raw-data centers hosted by BNL, DESY, GridKa, IN2P3, INFN, 
and the University of Victoria.

\bibliographystyle{JHEP}
\bibliography{references}

\end{document}